\documentclass{article}
\usepackage{graphicx}
\usepackage[a4paper, margin={1in}]{geometry}

\usepackage[labelfont=bf, font=bf]{caption}
\usepackage[singlelinecheck=false]{subcaption}
\usepackage{dcolumn}
\usepackage{bm}
\usepackage[colorlinks, urlcolor = {blue}]{hyperref}
\usepackage{color}
\usepackage[normalem]{ulem} 
\usepackage[english]{babel}
\usepackage{amsmath,amssymb}
\usepackage{indentfirst}
\usepackage{wrapfig}
\usepackage{url}
\usepackage{braket}
\usepackage{siunitx}
\usepackage{xfrac}
\usepackage{authblk}
\usepackage[colorinlistoftodos]{todonotes}
\usepackage{lineno}
\begin{document}
\title{Anomalous weak values in a generalized Mach--Zehnder interferometer extracted directly from intensity measurements}
\author[1,2]{Ismaele V. Masiello}
\author[1,3]{Hartmut Lemmel}
\author[1]{Andreas Dvorak}
\author[1]{Stephan Sponar}
\author[1,2]{Yuji Hasegawa}
\affil[1]{\small Atominstitut, TU Wien, Stadionallee 2, 1020 Vienna, Austria}
\affil[2]{\small Vienna Center for Quantum Science and Technology (VCQ), Austria}
\affil[3]{\small Institut Laue-Langevin, 71 avenue des Martyrs, 38000 Grenoble, France}
\date{\normalsize\today}
\maketitle
\small{Corresponding author: Ismaele V. Masiello (ismaele.masiello@tuwien.ac.at)} 

\begin{abstract}
Weak values provide a powerful framework for characterizing quantum systems. Their experimental extraction conventionally relies on weak conditioned von Neumann measurements, involving weak interactions and meter states that increase experimental complexity and often limit measurement efficiency. Here we introduce a method to fully characterize path weak-values in a generalized Mach--Zehnder interferometer employing neither meter states nor weak interactions. We experimentally demonstrate the technique in matter-wave interferometry. We identify anomalous weak values and, equivalently, negative quasiprobability distributions, which reflect the nonclassical behavior of the quantum system. The approach relies uniquely on intensity measurements at the output ports of the interferometer combined with controlled relative phase shifts between the paths. The absence of meter states enables considerable simplification of the setup and shorter measurement times, while preserving full access to weak values with comparable or increased accuracy. The scheme is directly applicable to a broad class of experiments involving two-level quantum systems.
\end{abstract}

\section{Introduction}\label{sec:Intro}
In the study of the physical properties of a quantum system between a pre-selected and a post-selected ensemble, i.e., between its initial state $\ket{\psi_\mathrm{in}}$ and final state $\ket{\psi_\mathrm{fi}}$, a measurement outcome known as the weak value arises in the context of weak measurements of an observable $\hat{A}$\,\cite{aharonov1988}. The weak value is defined as
\begin{align}
\frac{\braket{\psi_\mathrm{fi}|\hat{A}|\psi_\mathrm{in}}}{\braket{\psi_\mathrm{fi}|\psi_\mathrm{in}}}
\end{align}
and, unlike the expectation value, this quantity can be complex and may lie outside the eigenvalue range of the observable. Since the first implementation of a weak measurement\,\cite{ritchie1991}, the physical relevance of weak values has been confirmed across a range of impactful experiments. For instance, weak-value amplification enabled the observation of extremely small pointer's shifts, leading to the first measurement of the spin Hall effect of light\,\cite{hosten2008}. Weak values are also employed for the direct measurement of the wave function\,\cite{lundeen2011} and for probing nonlocal observables in Hardy’s paradox\,\cite{lundeen2009}. Experiments beyond photonic systems, such as the observation of the quantum Cheshire cat\,\cite{denkmayr2014, wagner2023, danner2024} and the observation of the quantum pigeonhole effect\,\cite{waegell2017}, both performed in neutron interferometry, further affirm the versatility of these techniques. The interpretation of weak values remains open to discussion\,\cite{dressel2015}. Nevertheless, they were found to play a role across several fundamental theoretical studies of quantum mechanics, including: quantum paradoxes\,\cite{aharonov2002}, uncertainty relations\,\cite{hall2004}, quasiprobability distributions\,\cite{ozawa2011, dressel2015, arvidsson-shukur2024}, nonclassicality\,\cite{spekkens2005, pusey2014,kunjwal2019},  and more\,\cite{dressel2010,dressel2012}. These far-reaching implications confirm that weak values constitute an extremely valuable extension of the standard measurement framework when one seeks to characterize quantum systems between preparation and post-selection, both conceptually and experimentally.

In order to exploit the advantages of weak values experimentally, effective and accurate methods of extraction are essential. Conventionally, weak values are obtained from weak measurements in which a meter system interacts weakly with the target system, followed by a readout of the meter\,\cite{hosten2008, lundeen2011, lundeen2009}. More recently, alternative schemes departing from this standard approach have been implemented. These include employing a classical parameter instead of a meter state while still maintaining a weak interaction\,\cite{masiello2025}, as well as using strong interaction while still retaining a meter state\,\cite{denkmayr2018}. The employment of meter states or the requirement of weak interactions can increase the experimental complexity. The manipulation of additional degrees of freedom increases the amount of required experimental resources, e.g., additional instrumentation and calibration procedures, while weak interactions typically necessitate longer measurement times due to the limited information gained per detection event.

In this paper, we propose a method to extract the complete path weak-values in a generalized Mach--Zehnder interferometer, requiring neither meter states nor weak interactions. The approach relies only on intensity measurements at the output ports of the interferometer combined with a controlled relative phase shift between the paths. The experimental verification is performed using a neutron Mach--Zehnder interferometer, enabling the study of the most fundamental element of quantum mechanics---the two-level quantum system---under conditions that challenge classical models: a single massive spin-1/2 particle in a superposition of spatially separated paths. Compared with typical photonic implementations, neutron interferometry is less susceptible to classical interpretations of the results. The setup requires only a small number of commonly used optical elements: two phase shifters, one for state preparation and the other for state manipulation, an absorber to control the relative path intensities, and a beam blocker. We further show that the same result can be obtained using a single phase shifter, reducing the number of required elements. Compared with previous neutron interferometer experiments extracting weak values using meter states or weak interactions\,\cite{masiello2025,denkmayr2018,danner2024_1,dvorak2025}, our method achieves equal or higher accuracy while substantially reducing the number of optical elements and significantly shortening the measurement time. Our results confirm the presence of anomalous weak values, i.e., weak values whose real part exceeds the eigenvalue range, which provide insight into the nonclassical behavior of the quantum system \,\cite{pusey2014, dressel2015, kunjwal2019}.

Neutron interferometry offers several advantages, such as macroscopic beam separation, individual control of the sub-beams, and long interaction and coherence times at room temperature and ambient pressure\,\cite{klepp2014, rauch2015, sponar2021, danner2023}. The monolithic crystal structure of the interferometer provides stable experimental conditions, enabling interference contrast that can exceed 90\%. Due to the fermionic nature of the neutron, the observed interference is intrinsically a single-particle phenomenon. The possibility to control additional degrees of freedom beyond the path states, such as spin and energy, makes neutron interferometry suitable for the investigation of nonclassical phenomena such as Bell-like inequalities and entanglement\,\cite{hasegawa2003, durstberger-rennhofer2011}. For these reasons, neutron interferometry has played a central role in experimental tests of fundamental aspects of quantum mechanics. The proposed scheme is not restricted to the path observables and post-selections of a generalized neutron Mach--Zehnder interferometer. The derivation is independent of the specific choice of two-level quantum system, observables, and post-selections, and therefore applicable to any quantum systems in which similar measurements and controlled phase shifts can be implemented. This makes the presented method suitable for applications in a variety of experiments across different fields, for instance in atom-optics experiments using Rabi oscillations or atom interferometers\,\cite{hornberger2012}, as well as polarimeter experiments\,\cite{klepp2014}.
\section{Results and discussion}\label{sec:Res}
\subsection{Path weak-values in a generalized Mach--Zehnder interferometer configuration}\label{sec:Theory}
\begin{figure}[t]
\centering
\includegraphics[width=0.8\linewidth,trim={4.5cm 12.5cm 4.5cm 12.5cm},clip]{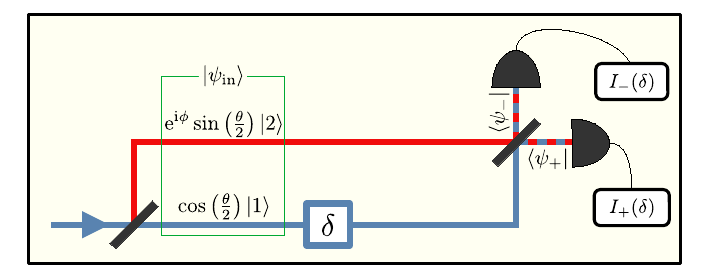}
\caption{\label{fig:inter_scheme}\bf{Generalized Mach--Zehnder interferometer configuration.}
\textnormal{The first beam splitter generates the superposition of path states $\ket{1}$ and $\ket{2}$, with relative phase $\phi$ and relative path intensities $\cos^2{\tfrac{\theta}{2}}$ and $\sin^2{\tfrac{\theta}{2}}$, which characterizes the initial state $\ket{\psi_\mathrm{in}}$. We refer to the case with $\theta=\tfrac{\pi}{2}$ (equal path intensities) as balanced configuration, otherwise it is unbalanced. A controlled phase shift $\delta$ is induced in one of the paths, in this case path 1. The last beam splitter, considered to be 50/50, projects on the exit beams $\ket{\psi_\pm}$. The intensities $I_{\pm}(\delta)$ is measured at the corresponding exit port.}}
\end{figure}

Consider the generalized Mach--Zehnder interferometer scheme shown in Fig.\,\ref{fig:inter_scheme}. No assumptions are made about the nature of the interfering quantum particle (e.g., photon, neutron, atom), and the first beam splitter is allowed to have arbitrary transmission and reflection coefficients. After passing through this beam splitter, the particle is prepared in a superposition of the path states $\ket{1}$ and $\ket{2}$, given by
\begin{align}
    \ket{\psi_\mathrm{in}}= \cos{\left(\tfrac{\theta}{2}\right)} 
    \ket{1} +
    \mathrm{e}^{\mathrm{i} \phi}\sin{\left(\tfrac{\theta}{2}\right)} \ket{2} \,,
    \label{eq:psi_j}
\end{align}
where $\cos^2{\left(\tfrac{\theta}{2}\right)}$ and $\sin^2{\left(\tfrac{\theta}{2}\right)}$ denote the relative intensities in path 1 and 2, respectively, and $\phi$ is the initial relative phase between the path states. The interferometer is said to operate in the balanced configuration when $\theta = \tfrac{\pi}{2}$, corresponding to the conventional 50/50 beam splitter. For any other value of $\theta$, the configuration is referred to as unbalanced. 

Next, we introduce an additional phase shift $\delta$ in path 1 transforming the initial state according to $\ket{\psi_\mathrm{in}}\rightarrow\mathrm{e}^{-\mathrm{i} \delta\hat{\Pi}_1} \ket{\psi_\mathrm{in}}$, where $\hat{\Pi}_j=\ket{j}\bra{j}$ is the path projector with $j=\{1,2\}$. Finally, at the second beam-splitter of the interferometer, the system is projected onto the final states
\begin{align}
    \ket{\psi_+} = \frac{\ket{1}+\ket{2}}{\sqrt{2}} \qquad \mathrm{and}\qquad
    \ket{\psi_-} = \frac{\ket{1}-\ket{2}}{\sqrt{2}} \, , \label{eq:+-}
\end{align}
corresponding to the two exit ports of the interferometer. This last beam splitter is assumed to be 50/50.
The intensities $I_{\pm}(\delta)$ measured at the output ports are proportional to the detection probability at the corresponding port
\begin{align}
\begin{aligned}
    I_{\pm}(\delta)=A\,\left|\braket{\psi_{\pm}|\mathrm{e}^{-\mathrm{i} \delta\hat{\Pi}_1} |\psi_\mathrm{in}} \right|^2 = A\left[ \frac{1}{2}\pm \frac{1}{2}\sin{\theta} \cos{(\phi+\delta)}\right] \, ,
\end{aligned} \label{eq:I+-}
\end{align}
where $A$ is a constant that depends on the setup characteristics and has the unit of count rate. This standard treatment of a generalized Mach--Zehnder interferometer predicts the complementary sinusoidal oscillations of the intensities measured at the two outgoing beams. For fixed relative path intensities, defined by $\theta$, and initial phase shift $\phi$, these oscillations depend on the phase $\delta$ and the obtained interference pattern is generally referred to as interferogram. Note that the amplitude of the interferogram is not proportional to the product of the relative path intensities $\cos^2{\left(\tfrac{\theta}{2}\right)}\sin^2{\left(\tfrac{\theta}{2}\right)}=\tfrac{1}{4}\sin^2{\theta}$, but on its square root. This distinction is key in differentiating between particles being in a quantum superposition of both paths and particles that are deterministically traveling through either one path or the other with a given probability\,\cite{summhammer1987}.

The path weak-values $w_{\pm,j}$, characterized by the initial state $\ket{\psi_\mathrm{in}}$, the observable $\hat{\Pi}_j$ and the final states $\ket{\psi_{\pm}}$, 
is a complex quantity defined as
\begin{align}
\begin{aligned}
    w_{\pm,j} = \frac{\braket{\psi_{\pm}|\hat{\Pi}_j|\psi_\mathrm{in}}}{\braket{\psi_{\pm}|\psi_\mathrm{in}}} =  w^\mathrm{R}_{\pm,j}+\mathrm{i}\, w^\mathrm{I}_{\pm,j} \, , \label{eq:wv}
\end{aligned}
\end{align}
where the terms $w^\mathrm{R}_{\pm,j}$ and $w^\mathrm{I}_{\pm,j}$ are, respectively, their real and imaginary part. Weak values are usually described in terms of weak conditioned von Neumann measurements, with a meter state that is first weakly coupled to the target observable and then measured to extract the components of the weak value\,\cite{aharonov1988, jozsa2007, hariri2019, dressel2014}. In our treatment, we do not make use of any meter state and the weak values can be extracted directly from measurements of the intensities of the outgoing beams, which depend on the phase shift parameter $\delta$ according to Eq.\,\eqref{eq:I+-}. In fact, the intensities $I_{\pm}(\delta)$ can be rewritten as
\begin{align}
\begin{aligned}
    I_{\pm}(\delta)&=A\,\left|\braket{\psi_{\pm}|\left[1+\left(\mathrm{e}^{-\mathrm{i} \delta}-1\right)\hat{\Pi}_1\right] |\psi_\mathrm{in}} \right|^2  = A\,\left|\braket{\psi_{\pm}| \psi_\mathrm{in}}\left[1 + \left(\mathrm{e}^{-\mathrm{i} \delta} - 1\right)w_{\pm,1}\right] \right|^2 \\
    &= A\,\left|\braket{\psi_{\pm}| \psi_\mathrm{in}} \right|^2\left[1 + 2\left(\lvert w_{\pm,1} \rvert^2 - w^\mathrm{R}_{\pm,1}\right)\left(1- \cos{\delta} \right) + 2\,w^\mathrm{I}_{\pm,1}\, \sin{\delta} \right]\\
    &=A\,\left|\braket{\psi_{\pm}| \psi_\mathrm{in}} \right|^2\left[1 + 2\left( \lvert w_{\pm,2} \rvert^2 - w^\mathrm{R}_{\pm,2}\right)\left(1 - \cos{\delta} \right) - 2\,w^\mathrm{I}_{\pm,2}\, \sin{\delta} \right] \, ,
\end{aligned} \label{eq:I+-_wv12}
\end{align}
where the first line is obtained from the relation
\begin{align}
\begin{aligned}
    \mathrm{e}^{-\mathrm{i} \delta\hat{\Pi}_1} &= \mathrm{e}^{-\mathrm{i} \delta\hat{\Pi}_1}\left(\hat{\Pi}_1 + \hat{\Pi}_2\right) = \mathrm{e}^{-\mathrm{i} \delta}\hat{\Pi}_1 + \hat{\Pi}_2  = 1+\left(\mathrm{e}^{-\mathrm{i} \delta}-1\right)\hat{\Pi}_1 \, ,
\end{aligned} \label{eq:exp_proj}
\end{align}
and the last line is derived by using the identity $\hat{\Pi}_1=1-\hat{\Pi}_2$.
The relation in Eq.\,\eqref{eq:I+-_wv12} suggests the direct extraction of the path weak-values from the intensities $I_{\pm}(\delta)$.

\subsection{Determination of the weak values from the intensity of the outgoing beams}
\begin{figure}[t]
\centering
\includegraphics[width=0.8\linewidth,trim={4.5cm 12.5cm 4.5cm 12.5cm},clip]{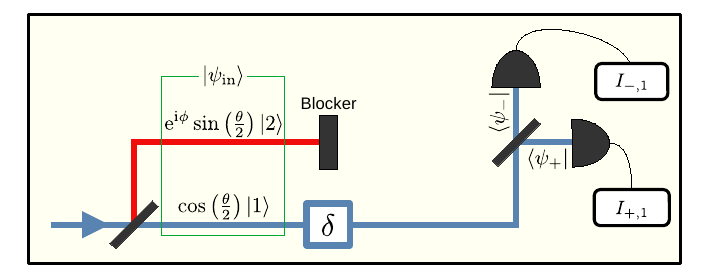}
\caption{\label{fig:blocker}\bf{Configuration for single-beam intensity measurement to be performed blocking each path.} \textnormal{A blocker is inserted in one of the paths, path 2 in the example depicted in the figure, in order to measure $I_{\pm,j}$}.}
\end{figure}
For simplicity of notation, we will focus only on the extraction of the weak values $w_{+,j}$, whose explicit expressions are given by
\begin{align}
    w_{+,1}= \frac{1}{1+ \tan{\left(\tfrac{\theta}{2}\right)}\mathrm{e}^{\mathrm{i} \phi}} \quad \mathrm{and} \quad
    w_{+,2} = \frac{1}{\mathrm{cotan}{\left(\tfrac{\theta}{2}\right)}\mathrm{e}^{-\mathrm{i} \phi}+1}\, .\label{eq:wv_expl}
\end{align}
The derivation of the real and imaginary components of $w_{-,j}$ can be obtained analogously. 

In the previous subsection, we expressed the outgoing intensity $I_{+}(\delta)$ of the generalized Mach--Zehnder interferometer in terms of the path weak-values. Each term in Eq.\,\eqref{eq:I+-_wv12} can be determined by appropriately tuning $\delta$ and exploiting the interferometer property $I_{+}(\delta+\pi)=I_{-}(\delta)$, namely
\begin{align}
     I_{+}(0) &= A\,\left|\braket{\psi_{+}| \psi_\mathrm{in}} \right|^2 \, , \label{eq:I_0}\\
    \frac{I_{-}(0)-I_{+}(0)}{4I_{+}(0)}  &= \lvert w_{+,1} \rvert^2 - w^\mathrm{R}_{+,1}  = \lvert w_{+,2} \rvert^2 - w^\mathrm{R}_{+,2}  \, , \label{eq:|w_1|2-w_1_re} \\
     \frac{I_{+}(\frac{\pi}{2})-I_{-}(\frac{\pi}{2})}{4 I_{+}(0)} &= w^\mathrm{I}_{+,1} = - w^\mathrm{I}_{+,2} \, . \label{eq:w_j_im}
\end{align}
The imaginary part $w^\mathrm{I}_{+,j}$ is directly obtained from Eq.\,\eqref{eq:w_j_im}.
The real part $w^\mathrm{R}_{+,j}$ enters quadratically in Eq.\,\eqref{eq:|w_1|2-w_1_re}, since $\lvert w_{+,j} \rvert^2={w^\mathrm{R}_{+,j}}^2+{w^\mathrm{I}_{+,j}}^2$. This leads to two possible solutions:
\begin{align}
\begin{aligned}
    w^R_{+,j} = \frac{1}{2} \pm \frac{1}{2}\sqrt{\frac{I_{-}(0)}{I_{+}(0)} -\left(2\,w^I_{+,j}\right)^2 }=\frac{1}{2}\pm\frac{1}{2} \sqrt{\frac{I_{-}(0)}{I_{+}(0)}-\left(\frac{I_{+}(\frac{\pi}{2})-I_{-}(\frac{\pi}{2})}{2 I_{+}(0)}\right)^2 }\, .
\end{aligned} \label{eq:w_j_re_sqrt}
\end{align}
The two solutions, corresponding to the different signs of the square root, must be assigned to either $w^R_{+,1}$ or $w^R_{+,2}$. At this stage, however, the assignment remains ambiguous. To resolve this ambiguity, we introduce an additional measurement of the relative path intensities. In doing so, we highlight the connection between the path intensities and the amplitude squared of the weak value $\left|w_{+,j} \right|^2$.

It is a straightforward task to resolve the relative intensity of the paths in the interferometer. A simple implementation consists of placing a blocker in one of the beams (e.g., path 2) as depicted in Fig.\,\ref{fig:blocker}. In such setup, the intensity $I_{+,1}$ of the unblocked path measured at the exit beam is calculated to be 
\begin{align}
\begin{aligned}
    I_{+,1} =A\, \left|\braket{\psi_{+}| \hat{\Pi}_1| \psi_\mathrm{in}} \right|^2 =A\,
    \left|\braket{\psi_{+}| \psi_\mathrm{in}} \right|^2\left|w_{+,1} \right|^2 =I_{+}(0) \left|w_{+,1} \right|^2=A\,\frac{\cos^2{\left(\tfrac{\theta}{2}\right)}}{2}  \, . \label{eq:block}
\end{aligned}
\end{align}
The measured intensity assumes a constant value (independent of the relative phase $\phi+\delta$), as no quantum interference effect occurs in a single-beam intensity measurement. The same procedure can be repeated blocking path 1 to obtain $I_{+,2}$. The relation in Eq.\,\eqref{eq:block} establishes a connection  between the relative path intensities measurement and the amplitude squared of the weak value. 

Here we emphasize that the measurement of $I_{+,j}$ enables two strategies for extracting the real part of the weak value. First, it can be combined with Eq.\,\eqref{eq:|w_1|2-w_1_re} to directly obtain
\begin{align}
    w^\mathrm{R}_{+,j}=\frac{I_{+,j}}{I_{+}(0)}+\frac{I_{+}(0)-I_{-}(0)}{4I_{+}(0)}\, . \label{eq:w_j_re}
\end{align}
Alternatively, Eq.\,\eqref{eq:w_j_re} can be used as a reference to fix the sign in Eq.\,\eqref{eq:w_j_re_sqrt}. The performances of these two methods are compared in Sec.\,\ref{sec:Met}.

A theoretical formulation of a method to extract weak values without weak measurement was presented by Johansen\,\cite{johansen2007_1}, in which the post-selection is performed prior the measurement of the target observable.  We should point out, however, that this approach cannot be applied to extract the weak values of the path projectors of the interferometer; the scheme in\,\cite{johansen2007_1} assumes the measurement of the path observables to be implemented \textit{after} the path recombination, i.e. on the outgoing beams $\ket{\psi_\pm}$. This situation is unfeasible in Mach--Zehnder interferometry, as the path degree of freedom cannot be accessed after the recombination at the last beam splitter.
\begin{figure}
\centering
\begin{subfigure}[t]{0.57\linewidth}
    \centering
\includegraphics[width=\linewidth,trim={0.5cm 8cm 0.5cm 8cm},clip]{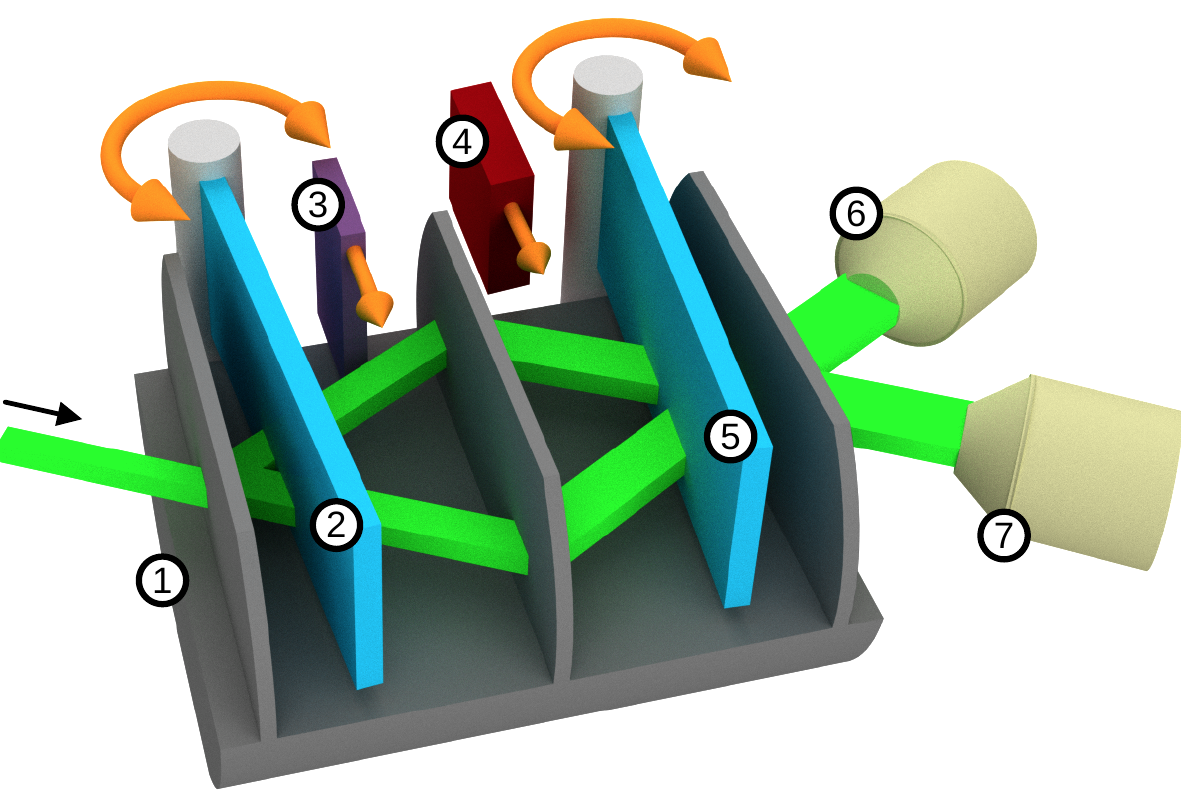}
\end{subfigure}
\begin{subfigure}[t]{0.42\linewidth}
    \centering
    \includegraphics[width=\linewidth]{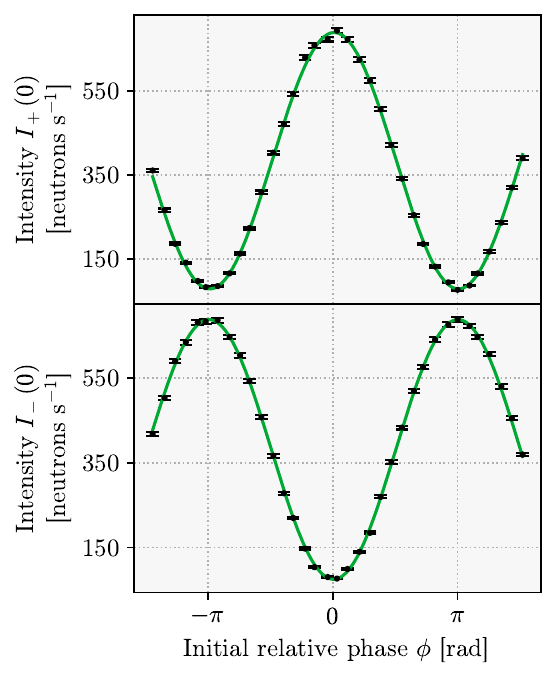}
\end{subfigure}
\caption{\label{fig:setup}\bf{3D representation of the neutron interferometer setup and typical sinusoidal intensity modulation depending on the relative phase $\phi$ of the initial state.} \textnormal{(On the left side) The neutron enters the silicon crystal interferometer \textcircled{1}. At the first plate, the neutron is prepared in a 50-50 superposition of the path states. A sapphire phase shifter plate \textcircled{2} placed just after the first plate is used to control the initial phase $\phi$. An absorber made of two indium foils \textcircled{3} is inserted to produce the unbalanced configuration (different path intensities) of the interferometer or removed to leave it in the balanced configuration (equal path intensities).  A Cadmium beam blocker \textcircled{4} can be inserted in path 2 to measure $I_{+,1}$ or in path 1 to measure $I_{+,2}$. A second phase shifter \textcircled{5} of the same kind of \textcircled{2} is placed  just before the beams-recombination to tune the phase shift $\delta$. Two $^3$He detectors are set to measure the intensity of the outgoing beams in the forward direction ($\ket{\psi_+}$-beam) \textcircled{7} and reflected direction ($\ket{\psi_-}$-beam) \textcircled{6}. (On the right side) Complementary intensity modulations of sinusoidal shape are recorded at the two detectors as a function of the initial relative phase $\phi$. Due to the inevitable off-set in the intensity of $\ket{\psi_-}$-beam, $I_-(0)$ is replaced by $I_+(\pi)$ (see text for more details). The data points are shown together with the theoretical predictions (continuous line), the error bars on the intensity measurements is taken to be the square root of the measured values (Gaussian uncertainty).}}
\end{figure}
\subsection{Setup and measurement}
The experiment was carried out at the neutron interferometry beamline S18 of the Institut Laue-Langevin (ILL)\,\cite{geppert2014}. A monochromatic neutron beam with wavelength $\lambda = 1.92$\,\AA\ and relative spread $\delta\lambda/\lambda \approx 0.02$ was employed. The setup comprised a triple-Laue silicon crystal interferometer together with standard neutron optical components; a schematic representation is shown in Fig.\,\ref{fig:setup}. At the first plate of the interferometer, each neutron is put in a superposition of two paths of equal intensity. The first phase shifter, a 2.4\,mm thick sapphire parallel-sided plate placed between the first and the second plate of the interferometer, is used to adjust the initial relative phase $\phi$. An absorber made of two indium foils of 1 mm and 0.5 mm thicknesses, with transmission of 48.3$\pm$0.2\% and 68.7$\pm$0.2\%, respectively, is placed in path 2 between the first phase shifter and the second plate to control the relative path intensities. When the absorber is inserted, the interferometer is in the unbalanced configuration with path intensities ratio of $\sin^2{\left(\tfrac{\theta}{2}\right)}/\cos^2{\left(\tfrac{\theta}{2}\right)}\approx 0.35$. When the absorber is removed, the interferometer is in the balanced configuration with path intensities ratio of $\sin^2{\left(\tfrac{\theta}{2}\right)}/\cos^2{\left(\tfrac{\theta}{2}\right)}\approx 1$. The first phase shifter and the indium absorber allow the preparation of the initial state $\ket{\psi_{\mathrm{in}}}$. 

A second phase shifter, with equal characteristics to the first one, is placed between the second and the third plate of the interferometer to tune the relative phase $\delta$. A beam blocker made of Cadmium can be inserted to realize the conditions of Fig.\,\ref{fig:blocker}, blocking either path 1 or path 2. The last plate transforms the two beams inside the interferometer into the two exit beams, which effectively decomposes the state $\mathrm{e}^{-\mathrm{i} \delta\hat{\Pi}_1} \ket{\psi_\mathrm{in}}$ into the final states $\ket{\psi_+}$ and $\ket{\psi_-}$. The intensity at the exit beams is measured using $^3$He detectors. We consider here only the intensity measurement performed on the $\ket{\psi_+}$-beam, since the $\ket{\psi_-}$-beam is inevitably accompanied by an offset due to the triple Laue diffraction configuration\,\cite{rauch2015}. In practice, the intensity of the $\ket{\psi_-}$-beam is obtained applying the phase shift $\delta\rightarrow\delta+\pi$, replacing effectively $I_-(0)$ and  $I_-(\tfrac{\pi}{2})$ with $I_+(\pi)$ and  $I_+(\tfrac{3\pi}{2})$, respectively.
\subsection{Experimental results}
\begin{figure}[t!]
\centering
\begin{subfigure}{0.49\linewidth}
    \centering
    \caption{\label{fig:meas_0}}
    \includegraphics[width=0.9\linewidth, trim={6cm 12.7cm 6cm 12.7cm},clip]{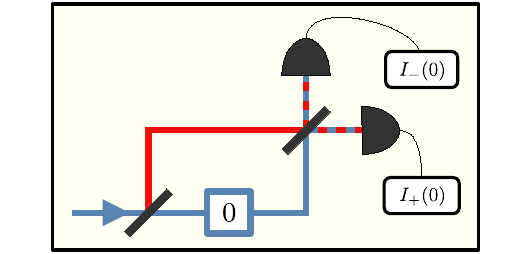}
    \includegraphics[width=0.7\linewidth]{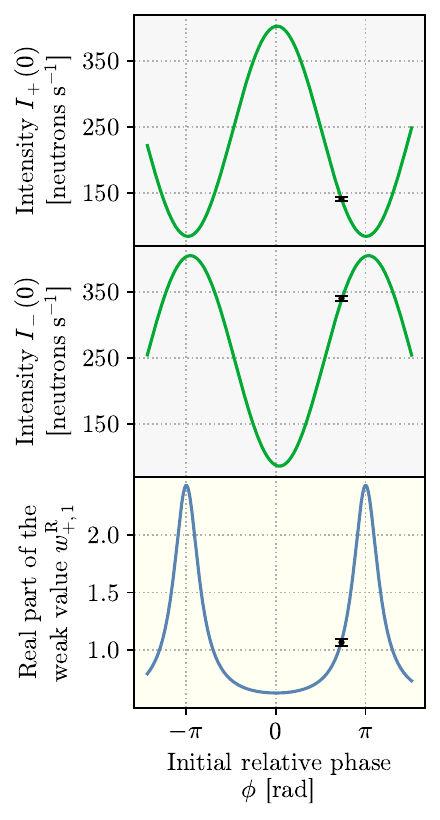}
\end{subfigure}
\begin{subfigure}{0.49\linewidth}
    \centering
    \caption{\label{fig:meas_pi2}}
    \includegraphics[width=0.9\linewidth, trim={6cm 12.7cm 6cm 12.7cm}, clip]{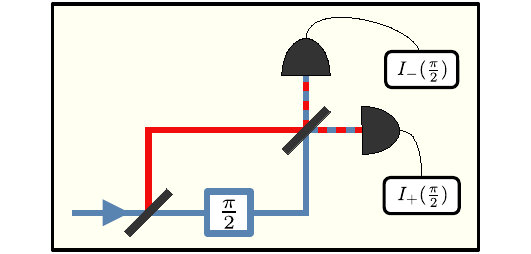}
    \includegraphics[width=0.7\linewidth]{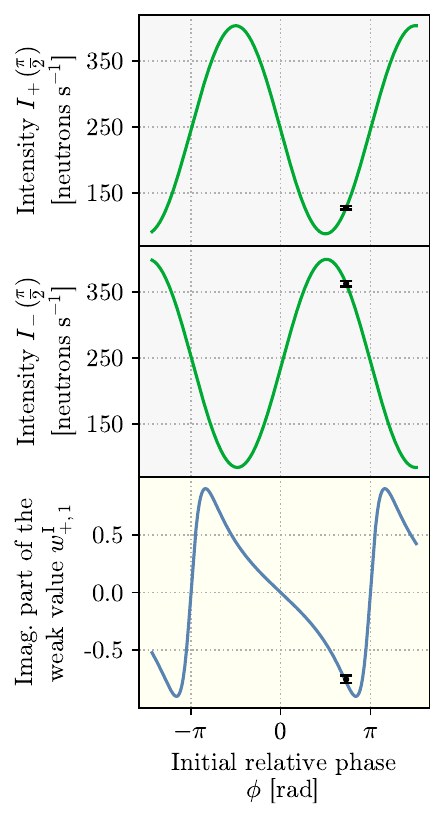}
\end{subfigure}
\caption{\label{fig:meas}\bf{Typical measurement procedure to extract (a) the real and (b) the imaginary part of the weak value, together with their results.} \textnormal{The initial state $\ket{\psi_\mathrm{in}}$ is generated with relative phase $\phi\approx3\pi/4$ and path intensities ratio $\sin^2{\left(\tfrac{\theta}{2}\right)}/\cos^2{\left(\tfrac{\theta}{2}\right)}\approx 0.35$. (a) The two intensities $I_\pm(\delta=0)$ are measured to extract the real part of the weak value, together with a measurement of $I_{+,1}$. (b) The two intensities $I_\pm(\delta=\tfrac{\pi}{2})$ are measured to extract the imaginary part of the weak value, together with $I_+(\delta=0)$. All results are obtained from the intensity measurements together with the data correction described in Sec.\,\ref{sec:Met}. The data points are shown together with the theoretical predictions (continuous line).}}
\end{figure}
\begin{figure}[t!]
\centering
\begin{subfigure}{0.495\linewidth}
\centering
\caption{\label{subfig:results real}}
\includegraphics[width=\linewidth]{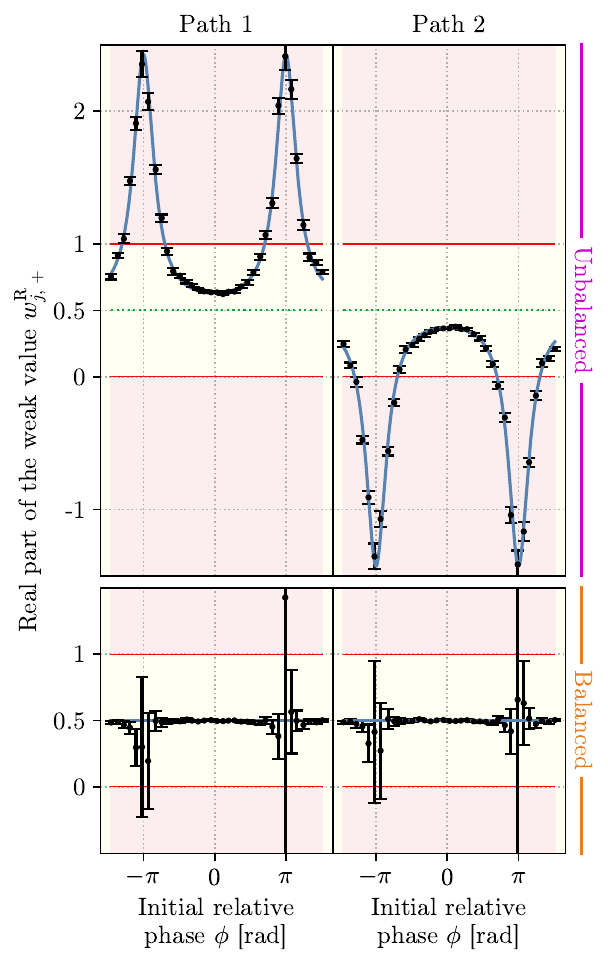}
\end{subfigure}
\begin{subfigure}{0.495\linewidth}
\centering
\caption{\label{subfig:results imag}}  
\includegraphics[width=\linewidth]{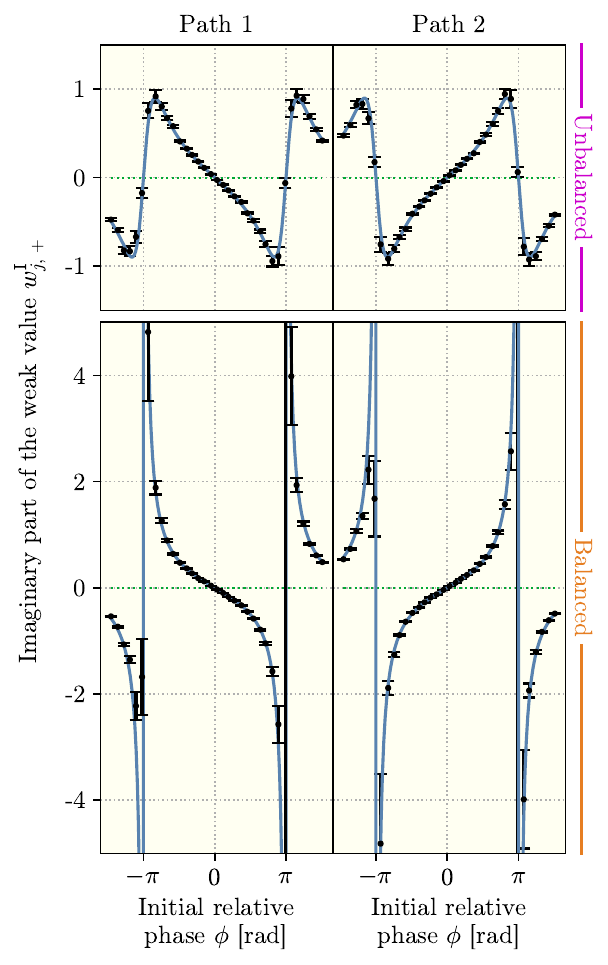}
\end{subfigure}
\caption{\label{fig:results}\bf{Resulting weak values (data points) together with the theoretical prediction (continuous line).} \textnormal{The results for the weak value are plotted as their (a) real and (b) imaginary part for the different values of the initial relative phase $\phi$. The top plots show the result for the unbalanced configuration the bottom plots show  the results for the balanced configuration. The results are in good agreement with the theory. The region in which the weak values become anomalous is highlighted in red. The green dotted line represents the symmetry axis of the weak-value components. The real part is symmetric about $0.5$, while the imaginary part is symmetric about $0$, confirming $w_{+,1}+w_{+,2}=1$.}}
\end{figure}
The interferograms performed during the calibration stage, similar to the ones shown in Fig.\,\ref{fig:setup}, exhibit good levels of phase homogeneity and high contrast of the interference fringes. The typical contrast is 64\% and 80\% for the unbalanced and balanced configuration, respectively. Once all the optical elements are in place, the measurement of $I_{+,1}$ is implemented by placing the blocker in path 2 and measuring the intensity. The same procedure is repeated blocking path 1 and measuring $I_{+,2}$. Afterwards, the blocker is removed and the measurements corresponding to $I_\pm(0)$ and $I_\pm(\tfrac{\pi}{2})$ are performed. An example of the latter measurements, performed on an initial state $\ket{\psi_\mathrm{in}}$ with initial relative phase $\phi\approx3\pi/4$ and path intensity ratio $\sin^2{\left(\tfrac{\theta}{2}\right)}/\cos^2{\left(\tfrac{\theta}{2}\right)}\approx 0.35$, is shown in Fig.\,\ref{fig:meas} together with the resulting weak values. The theoretical predictions are shown using a continuous line. It is confirmed here that the real part of the weak value can be extracted from a measurement of the outgoing beams of a generalized Mach--Zehnder interferometer at a single phase shifter position, together with the measurement of the single-beam intensity. Additionally, Fig.\,\ref{fig:meas} shows that a shift in $\delta$ is equivalent to a shift in $\phi$, in accordance with Eq.\,\eqref{eq:I+-}. Consequently, the experiment can be conducted employing a single phase shifter for the preparation and manipulation of the state.

The measured path weak-values $w_{+,j}$ are presented in Fig.\,\ref{fig:results}, together with their theoretical predictions. The results for $w_{-,j}$ are not shown, but can be obtained analogously. The real and imaginary components, $w^R_{+,j}$ and $w^I_{+,j}$, are displayed in Fig.\,\ref{subfig:results real} and Fig.\,\ref{subfig:results imag}, respectively. In each figure, the unbalanced configuration is presented in the upper two panels, while the balanced configuration appears in the lower two panels. The initial states are characterized by the initial relative phase $\phi$ and the relative path intensities $\cos^2{\left(\tfrac{\theta}{2}\right)}$ and $\sin^2{\left(\tfrac{\theta}{2}\right)}$; therefore, each value of $\phi$ corresponds to a distinct initial state, for a total of 35 initial states per configuration. The error bars are obtained via standard error propagation from the uncertainties in the measured intensities and in the correction parameters, the latter presented in detail in Sec.\,\ref{sec:Met}. 

The real part of the weak value for the unbalanced configuration, obtained according to Eq.\,\eqref{eq:w_j_re_sqrt}, is shown in the top panels of Fig.\,\ref{subfig:results real}. The region outside the eigenvalue range of the path projectors is highlighted in red. When the real components enter this region, the weak values become anomalous. Such anomalous weak values are equivalent to negative quasiprobability distributions and thus reflect nonclassical behavior\,\cite{dressel2015}. They also constitute proof of generalized contextuality\,\cite{spekkens2005, pusey2014, kunjwal2019}. The bottom panels of Fig.\,\ref{subfig:results real} display the real part of the weak value for the balanced configuration, extracted using Eq.\,\eqref{eq:w_j_re}. In this case, no anomalous weak values are observed; hence, the balanced configuration does not exhibit nonclassical behavior\,\cite{catani2023}. The experimental results for both configurations are in good agreement with theory. Larger error bars and increased fluctuations are visible in the balanced configuration in the vicinity of $\phi=\pm\pi$. This behavior originates from the term $I_+(0)$ approaching zero in the denominator of Eq.\,\eqref{eq:w_j_re}, making the result more sensitive to intensity fluctuations.

The results for the imaginary parts of the weak values are presented in Fig.\,\ref{subfig:results imag}. The results for both the balanced (top panels) and unbalanced (bottom panels) configurations are obtained from Eq.\,\eqref{eq:w_j_im}. The data show good agreement with the theoretical prediction. In the balanced configuration, the term $I_+(0)$ approaches zero in the denominator of Eq.\,\eqref{eq:w_j_im} and the imaginary part tends to diverge. Therefore, more noticeable fluctuations and enlarged error bars appear around $\phi=\pm\pi$, similarly to the real part in the same configuration.

In all graphs, the green dotted line represents the symmetry axis of the weak-value components: the real part is symmetric about $0.5$, while the imaginary part is symmetric about $0$. This highlights that the data confirm the property $w_{+,1}+w_{+,2}=1$, or, equivalently, $w^\mathrm{R}_{+,1} - \tfrac{1}{2}=-w^\mathrm{R}_{+,2} +\tfrac{1}{2}$ and $w^\mathrm{I}_{+,1}+w^\mathrm{I}_{+,2}=0$.
\subsection{Discussion}
Neutron interferometer experiments involving the manipulation of a meter state, conventionally being the spin degree of freedom, are subject to  substantial neutron losses and contrast reduction. The polarization, typically implemented using magnetic prisms, discards approximately half of the incident neutrons. Spin rotations, usually performed with magnetic coils inside the interferometer, require a reduced beam cross-section and introduce thermal disturbance which reduces the contrast. The spin analysis for thermal neutron interferometry, generally performed with a polarizing multi-layer array (``supermirror''), has a typical output efficiency of 10-20\% with relatively small beam cross section\,\cite{geppert2014}. The absence of spin manipulation in our experiment results in higher neutron count rate and enhanced visibility of the interference fringes, leading to shorter acquisition times while maintaining comparable or improved precision. For instance, either the top or the bottom panels of Fig.\,\ref{fig:results} can be acquired in about 1 hour of measurement time, constituting a reduction of one order of magnitude compared with similar neutron interferometer experiments\,\cite{masiello2025}.

Apart from the calibration procedure, only three measurement configurations (corresponding to five intensity measurements) are required to fully characterize the path weak-values in a Mach--Zehnder interferometer: $I_\pm(0)$, $I_\pm(\tfrac{\pi}{2})$ and $I_{+,1}$. Additionally, if the aim is to witness anomalous weak values, only two configurations (four intensity measurements) are sufficient, namely $I_\pm(0)$ and $I_\pm(\tfrac{\pi}{2})$. In this case, Eq.\,\eqref{eq:w_j_re_sqrt} can reveal the presence of anomalous weak values without fixing the sign of the square root, since both solutions would lie outside the eigenvalue range of the path projectors. 

For the chosen observables and post-selection, the real part of the weak values $w^\mathrm{R}_{\pm,1}$ depends only on the expectation values of the Pauli operators $\hat{\sigma}_x$ and $\hat{\sigma}_z$ for the initial state $\ket{\psi_\mathrm{in}}$, with the path states being on the quantization axis $z$, while the imaginary part $w^\mathrm{I}_{\pm,1}$ depends only on the expectation values of $\hat{\sigma}_x$ and $\hat{\sigma}_y$. Consistently, it is sufficient to measure $I_{+,j}$ (a measurement in the path basis corresponding to the $z$-axis of the Bloch sphere) and at the output ports without an additional phase shift $I_\pm(0)$ (corresponding to the $x$-axis), to fully characterize the real part of the weak value. The imaginary part is extracted from output measurements with and without a phase shift of $\pi/2$, $I_\pm(0)$ and $I_\pm(\pi/2)$, corresponding to the $x$- and $y$-axes, respectively. Remarkably, Eq.\,\eqref{eq:w_j_re_sqrt} shows that anomalous weak values, a feature linked to the real part, can be inferred from measurements of $\hat{\sigma}_x$ and $\hat{\sigma}_y$ alone.

The results presented in this paper are not restricted to Mach--Zehnder interferometry, neither to the specific choice of path projectors $\hat{\Pi}_j$ nor to the post-selections $\ket{\psi_\pm}$. In fact, the derivation of Eq.\,\ref{eq:I+-_wv12} is independent of both the projectors employed and the post-selection performed. For example, consider neutron polarimeter experiments, which operate on the neutron’s spin degree of freedom\,\cite{klepp2014}. In such setups, phase shifts can be implemented about arbitrary axes and measurements can be performed along arbitrary directions. Consequently, the weak value of any spin observable---each of which can be decomposed into a sum of projectors---can, in principle, be extracted for any chosen post-selection. Furthermore, the authors of the present paper are currently working on the extension of the method to n-level systems. Similarly to the method introduced by Johansen\,\cite{johansen2007}, our approach can also be used to extract the Kirkwood–Dirac quasiprobability distribution. Specifically, the terms correspond to $w_{\pm,j}\,I_{\pm}(0)/A$\,\cite{arvidsson-shukur2024}.

Last, we discuss here another relation between the extraction of weak values and the four intensities $I_\pm(0)$ and $I_\pm(\frac{\pi}{2})$. By using the Taylor series, the intensity of the outgoing beams $I_{\pm}(\delta)$ in Eq.\,\eqref{eq:I+-} can be expanded to be $I_{\pm}(\delta+\Delta\delta) = I_{\pm}(\delta) + \Delta\delta \cdot I'_{\pm}(\delta) +O(\Delta\delta^2)$ where the derivative $I'_{\pm}(\delta)$ of $I_{\pm}(\delta)$ is calculated to be
$I'_{\pm}(\delta) = 2 \left|\braket{\psi_{\pm}| \psi_\mathrm{in}} \right|^2\left[\left(\lvert w_{\pm,1} \rvert^2 - w^\mathrm{R}_{\pm,1}\right)\sin{\delta} + \,w^\mathrm{I}_{\pm,1}\, \cos{\delta} \right]$.
Considering the intensity difference in the case of small phase-shift $\Delta\delta$ and tuning the phase shift $\delta=0,\pi/2$, we get, 
\begin{align}
\frac{I_{\pm}(0+\Delta\delta) - I_{\pm}(0)}{\Delta\delta} &\sim  I'_{\pm}(0) = 2A\,\left|\braket{\psi_{\pm}| \psi_\mathrm{in}} \right|^2 w^\mathrm{I}_{\pm,1},
 \\
\frac{I_{\pm}(\pi/2+\Delta\delta) - I_{\pm}(\pi/2)}{\Delta\delta} &\sim  I'_{\pm}(\pi/2) = 2A\,\left|\braket{\psi_{\pm}| \psi_\mathrm{in}} \right|^2 \left(\lvert w_{\pm,1} \rvert^2 - w^\mathrm{R}_{\pm,1}\right).
\end{align}
In the present case, owing to the sinusoidal dependence of $I_{\pm}(\delta)$ on $\delta$ (see Eq.\,\eqref{eq:+-}), the differential coefficient $I'_{\pm}(\delta)$ is given directly by $\left[I_{\pm}(\delta+\pi/2)-I_{\pm}(\delta-\pi/2)\right]/2$, so that terms such as $w^\mathrm{I}_{\pm,j}$ and $\left(\lvert w_{\pm,j} \rvert^2 - w^\mathrm{R}_{\pm,j}\right)$ $(j=1,2)$ are accessed through the four intensities of $I_\pm(\delta)$ (see, Eqs.\,(\ref{eq:|w_1|2-w_1_re}-\ref{eq:w_j_im})). Note that, $I_\pm(\delta)$ and $\delta$ can be regarded as the meter-shift function and the interaction strength, respectively, in a conventional weak measurement scheme. We suggest that 
the meter-shift divided by the interaction strength, as in the first terms of the above equalities, appears here as well as in the determination of weak values in the standard weak measurement scheme\,\cite{dressel2014}; it is shown to be directly connected to the differential coefficient of the meter-shift function.

\section{Methods}\label{sec:Met}
\subsection{Weak value extraction from data}\label{subsec:WvExtrData}
\begin{figure}[t]
\begin{subfigure}{0.495\linewidth}
\centering
\caption{\label{fig:wrong_re_unb}}
\includegraphics[width=\linewidth]{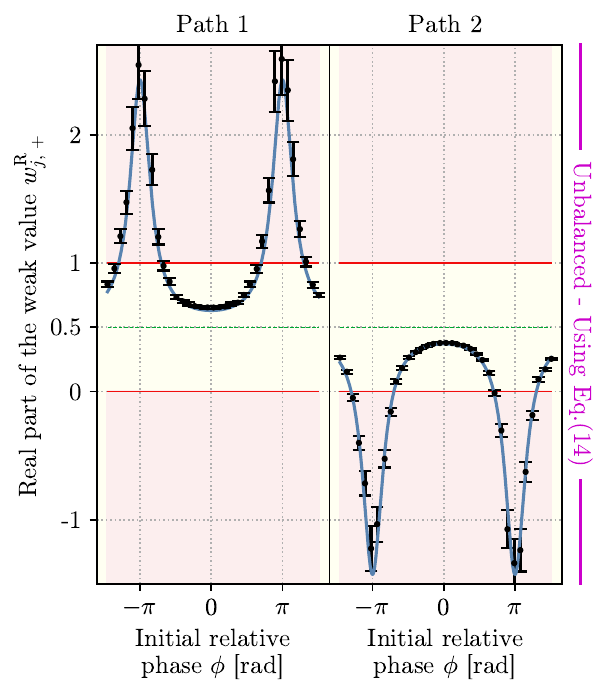}
\end{subfigure}
\begin{subfigure}{0.495\linewidth}
\centering
\caption{\label{fig:wrong_re_bal}}
\includegraphics[width=\linewidth]{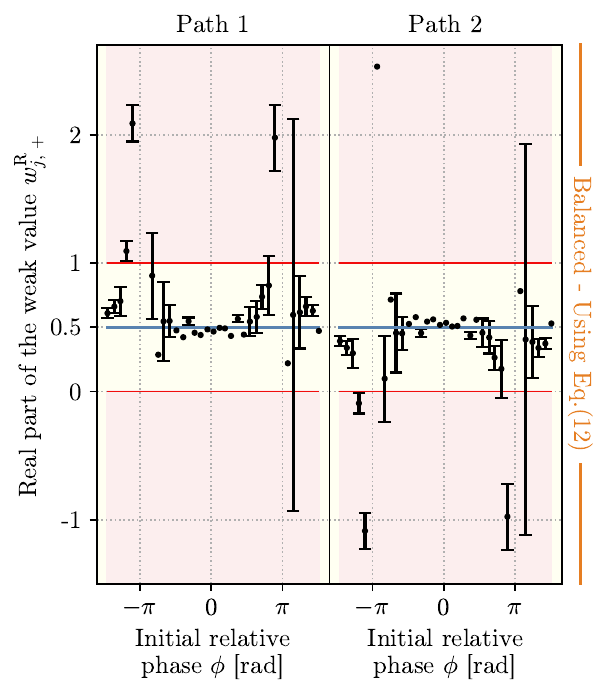}
\end{subfigure}
\caption{\label{fig:wrong_re}\bf{Example of an alternative choice of extraction method for the real part of the weak values.} \textnormal{The results shown in Fig.\,\ref{subfig:results real} are obtained using Eq.\,\eqref{eq:w_j_re_sqrt}  for the unbalanced and Eq.\,\eqref{eq:w_j_re} for the balanced configuration to get optimal results. Here we invert our choice for comparison. (a) In the unbalanced configuration, the use of Eq.\,\eqref{eq:w_j_re} introduces a systematic deviation from the theoretical prediction due to the manual insertion of the beam blocker. 
(b) In the balanced configuration, Eq.\,\eqref{eq:w_j_re_sqrt} produces enhanced fluctuations and larger error bars due to the instability of the square-root argument near equal path intensities.}}
\end{figure}
The description presented in Sec.\,\ref{sec:Theory} assumes idealized experimental conditions. In order to extract the weak values from the experimental data, the model has to account for a reduction in contrast of the interference fringes due to phase inhomogeneities caused by real-life imperfections\,\cite{klepp2014}. The experimentally detected intensity $I^{\mathrm{exp}}_{+}(\delta)$ is modeled as 
\begin{align}
\begin{aligned}
        I^{\mathrm{exp}}_{+}(\delta) = A\left[\frac{1}{2}+ \frac{C}{2} \sin\theta \cos{(\delta+\phi)}\right] = A\left[C\,I_+(\delta)+\frac{1-C}{2}\right]\, ,
    \label{eq:I_meas}
\end{aligned}
\end{align}
where $A$ is a constant with units of neutron count rate and $C$ is the contrast reduction factor. Analogously, we define the experimentally measured intensity with one of the path blocked (Fig.\,\ref{fig:blocker}) as $I^{\mathrm{exp}}_{+,j}=A\,I_{+,j}$. Here, no contrast reduction appears since a single beam does not produce interference effects.

The parameters $A$ and $C$ are estimated from interferograms (Fig.\,\ref{fig:setup}) as twice the average intensity and as the oscillation amplitude divided by $A \sin\theta$, respectively. The ideal oscillation amplitude $\sin\theta$ is obtained from the intensities $I^{\mathrm{exp}}_{+,j}$ as\,\cite{summhammer1987}
\begin{align}
    \sin\theta = 2\frac{\sqrt{I^{\mathrm{exp}}_{+,1}I^{\mathrm{exp}}_{+,2}}}{I^{\mathrm{exp}}_{+,1}+I^{\mathrm{exp}}_{+,2}}= 2\frac{\sqrt{I_{+,1}I_{+,2}}}{I_{+,1}+I_{+,2}} \,. 
\end{align}
The corrected intensities are defined as
\begin{align}
\begin{aligned}
    I^{\mathrm{corr}}_{+}(\delta) =  I^{\mathrm{exp}}_{+}(\delta)-A \frac{1-C}{2} = AC\,I_{+}(\delta) \quad \mathrm{and} \quad I^{\mathrm{corr}}_{+,j}=AC\,I^{\mathrm{exp}}_{+,j} = AC\,I_{+,j}\, ,\label{eq:I_meas_corr}  
\end{aligned}
\end{align}
the relative path intensities measurements are multiplied by the contrast reduction factor for consistency, in order to extract the path weak-values $w_{\pm,j}$ by substituting $I^{\mathrm{corr}}_{+}(\delta)$ and $I^{\mathrm{corr}}_{+,j}$ in Eqs.\,\eqref{eq:w_j_im}, \eqref{eq:w_j_re}, and \eqref{eq:w_j_re_sqrt}. Note that no correction would be needed in a setup with negligible contrast reduction ($C\approx1$).

In the present experiment, the beam blocker is inserted manually while the phase shifters are controlled remotely.
Placing the beam blocker by hand can alter the environmental conditions, e.g., altering the room temperature or generating of air flows, affecting the average neutron count rate and the contrast reduction factor. In this situation, it can be advantageous to extract the real part of the weak value using Eq.\,\eqref{eq:w_j_re_sqrt}: the measurements of $I^{\mathrm{exp}}_{\pm}(0)$ and $I^{\mathrm{exp}}_{\pm}(\tfrac{\pi}{2})$ can be implemented without accessing the interferometer room, minimizing the fluctuations of $A$ and $C$. Nevertheless, a measurement of $I^{\mathrm{exp}}_{+,j}$ is required to fix the sign of the square root, using Eq.\,\eqref{eq:w_j_re} as a reference. However, it is not always convenient to use Eq.\,\eqref{eq:w_j_re_sqrt}. Approaching equal path intensities, the imaginary part of the weak value diverges near $\phi=\pm\pi$, making the argument of the square root not stable. In this regime, Eq.\,\eqref{eq:w_j_re} provides a method that is not subject to the divergence of the imaginary part. 

An example of an alternative choice of extraction method is shown in Fig.\,\ref{fig:wrong_re}. In Fig.\,\ref{fig:wrong_re_unb}, the real part of the weak value in the unbalanced configuration is extracted using Eq.\,\eqref{eq:w_j_re}, resulting in a systematic deviation from the theoretical curve due to manual insertion of the beam blocker. 
Conversely, in Fig.\,\ref{fig:wrong_re_bal}, the use of Eq.\,\eqref{eq:w_j_re_sqrt} in the balanced configuration leads to enhanced fluctuations and enlarged error bars, reflecting the instability of the square-root argument.
\section*{Data availability}
The data that support the findings of this study are openly available at the following URL/DOI: \url{http://doi.ill.fr/10.5291/ILL-DATA.CRG-3125}.
\bibliographystyle{naturemag}
\bibliography{references}

@PREAMBLE{
 "\providecommand{\noopsort}[1]{}" 
 # "\providecommand{\singleletter}[1]{#1}%" 
}

@article{aharonov1988,
  title = {How the result of a measurement of a component of the spin of a spin-1/2 particle can turn out to be 100},
  author = {Aharonov, Yakir and Albert, David Z. and Vaidman, Lev},
  journal = {Phys. Rev. Lett.},
  volume = {60},
  issue = {14},
  pages = {1351--1354},
  numpages = {0},
  year = {1988},
  month = {Apr},
  publisher = {American Physical Society},
  doi = {10.1103/PhysRevLett.60.1351},
  url = {https://link.aps.org/doi/10.1103/PhysRevLett.60.1351}
}

@article{hall2004,
  title = {Prior information: How to circumvent the standard joint-measurement uncertainty relation},
  author = {Hall, Michael J. W.},
  journal = {Phys. Rev. A},
  volume = {69},
  issue = {5},
  pages = {052113},
  numpages = {12},
  year = {2004},
  month = {May},
  publisher = {American Physical Society},
  doi = {10.1103/PhysRevA.69.052113},
  url = {https://link.aps.org/doi/10.1103/PhysRevA.69.052113}
}

@article{dressel2015,
  title = {Weak values as interference phenomena},
  author = {Dressel, Justin},
  journal = {Phys. Rev. A},
  volume = {91},
  issue = {3},
  pages = {032116},
  numpages = {14},
  year = {2015},
  month = {Mar},
  publisher = {American Physical Society},
  doi = {10.1103/PhysRevA.91.032116},
  url = {https://link.aps.org/doi/10.1103/PhysRevA.91.032116}
}

@article{aharonov2002,
title = {Revisiting Hardy's paradox: counterfactual statements, real measurements, entanglement and weak values},
journal = {Phys. Lett. A},
volume = {301},
number = {3},
pages = {130-138},
year = {2002},
issn = {0375-9601},
doi = {https://doi.org/10.1016/S0375-9601(02)00986-6},
url = {https://www.sciencedirect.com/science/article/pii/S0375960102009866},
author = {Yakir Aharonov and Alonso Botero and Sandu Popescu and Benni Reznik and Jeff Tollaksen}
}

@article{dressel2014,
  title = {Colloquium: Understanding quantum weak values: Basics and applications},
  author = {Dressel, Justin and Malik, Mehul and Miatto, Filippo M. and Jordan, Andrew N. and Boyd, Robert W.},
  journal = {Rev. Mod. Phys.},
  volume = {86},
  issue = {1},
  pages = {307--316},
  numpages = {10},
  year = {2014},
  month = {Mar},
  publisher = {American Physical Society},
  doi = {10.1103/RevModPhys.86.307},
  url = {https://link.aps.org/doi/10.1103/RevModPhys.86.307}
}

@article{dressel2012,
   title={Significance of the imaginary part of the weak value},
   volume={85},
   ISSN={1094-1622},
   url={http://dx.doi.org/10.1103/PhysRevA.85.012107},
   DOI={10.1103/physreva.85.012107},
   number={1},
   journal={Phys. Rev. A},
   publisher={American Physical Society (APS)},
   author={Dressel, J. and Jordan, A. N.},
   year={2012},
   month=jan }

@article{dressel2010,
   title={Contextual Values of Observables in Quantum Measurements},
   volume={104},
   ISSN={1079-7114},
   url={http://dx.doi.org/10.1103/PhysRevLett.104.240401},
   DOI={10.1103/physrevlett.104.240401},
   number={24},
   journal={Phys. Rev. Lett.},
   publisher={American Physical Society (APS)},
   author={Dressel, J. and Agarwal, S. and Jordan, A. N.},
   year={2010},
   month=jun }

@article{jozsa2007,
   title={Complex weak values in quantum measurement},
   volume={76},
   ISSN={1094-1622},
   url={http://dx.doi.org/10.1103/PhysRevA.76.044103},
   DOI={10.1103/physreva.76.044103},
   number={4},
   journal={Phys. Rev. A},
   publisher={American Physical Society (APS)},
   author={Jozsa, Richard},
   year={2007},
   month=oct }

@article{lundeen2009,
   title={Experimental Joint Weak Measurement on a Photon Pair as a Probe of Hardy’s Paradox},
   volume={102},
   ISSN={1079-7114},
   url={http://dx.doi.org/10.1103/PhysRevLett.102.020404},
   DOI={10.1103/physrevlett.102.020404},
   number={2},
   journal={Phys. Rev. Lett.},
   publisher={American Physical Society (APS)},
   author={Lundeen, J. S. and Steinberg, A. M.},
   year={2009},
   month=jan }

@article{denkmayr2014,
  title={Observation of a quantum Cheshire Cat in a matter-wave interferometer experiment},
  author={Denkmayr, Tobias and Geppert, Hermann and Sponar, Stephan and Lemmel, Hartmut and Matzkin, Alexandre and Tollaksen, Jeff and Hasegawa, Yuji},
  journal={Nat. Commun.},
  volume={5},
  number={1},
  pages={4492},
  year={2014},
  publisher={Nature Publishing Group UK London}
}

@article{denkmayr2018,
title = {Weak values from strong interactions in neutron interferometry},
journal = {Phys. B Condensed Matter.},
volume = {551},
pages = {339-346},
year = {2018},
issn = {0921-4526},
doi = {https://doi.org/10.1016/j.physb.2018.04.014},
url = {https://www.sciencedirect.com/science/article/pii/S0921452618302722},
author = {Tobias Denkmayr and Justin Dressel and Hermann Geppert-Kleinrath and Yuji Hasegawa and Stephan Sponar}
}

@article{danner2024,
  title={Three-path quantum Cheshire cat observed in neutron interferometry},
  author={Danner, Armin and Geerits, Niels and Lemmel, Hartmut and Wagner, Richard and Sponar, Stephan and Hasegawa, Yuji},
  journal={Commun. Phys.},
  volume={7},
  number={1},
  pages={14},
  year={2024},
  publisher={Nature Publishing Group UK London}
}

@inproceedings{ozawa2011,
   title={Universal Uncertainty Principle, Simultaneous Measurability, and Weak Values},
   ISSN={0094-243X},
   url={http://dx.doi.org/10.1063/1.3630147},
   DOI={10.1063/1.3630147},
   booktitle={AIP Conference Proceedings},
   publisher={AIP},
   author={Ozawa, Masanao and Ralph, Timothy and Lam, Ping Koy},
   year={2011} }

@article{hariri2019,
  title = {Experimental simultaneous readout of the real and imaginary parts of the weak value},
  author = {Hariri, A. and Curic, D. and Giner, L. and Lundeen, J. S.},
  journal = {Phys. Rev. A},
  volume = {100},
  issue = {3},
  pages = {032119},
  numpages = {7},
  year = {2019},
  month = {Sep},
  publisher = {American Physical Society},
  doi = {10.1103/PhysRevA.100.032119},
  url = {https://link.aps.org/doi/10.1103/PhysRevA.100.032119}
}

@article{masiello2025,
doi = {10.1088/1367-2630/adb175},
url = {https://dx.doi.org/10.1088/1367-2630/adb175},
year = {2025},
month = {feb},
publisher = {IOP Publishing},
volume = {27},
number = {2},
pages = {023017},
author = {Masiello, Ismaele V and Dvorak, Andreas and Lemmel, Hartmut and Danner, Armin and Hasegawa, Yuji},
title = {Simultaneous determination of two path weak-values with time-dependent phase manipulation in neutron interferometry},
journal = {New J. Phys.}
}

@article{geppert2014,
title = {Improvement of the polarized neutron interferometer setup demonstrating violation of a Bell-like inequality},
journal = {Nucl. Instrum. Methods Phys. Res. A},
volume = {763},
pages = {417-423},
year = {2014},
issn = {0168-9002},
doi = {https://doi.org/10.1016/j.nima.2014.06.080},
url = {https://www.sciencedirect.com/science/article/pii/S0168900214008341},
author = {H. Geppert and T. Denkmayr and S. Sponar and H. Lemmel and Y. Hasegawa}
}

@article{dvorak2025,
  title = {Tight qubit uncertainty relations studied through weak values in neutron interferometry},
  author = {Dvorak, Andreas and Masiello, Ismaele V. and Hasegawa, Yuji and Lemmel, Hartmut and Hofmann, Holger F. and Sponar, Stephan},
  journal = {Phys. Rev. Res.},
  volume = {7},
  issue = {4},
  pages = {043334},
  numpages = {9},
  year = {2025},
  month = {Dec},
  publisher = {American Physical Society},
  doi = {10.1103/pthn-81pm},
  url = {https://link.aps.org/doi/10.1103/pthn-81pm}
}

@article{johansen2007,
  title = {Quantum theory of successive projective measurements},
  author = {Johansen, Lars M.},
  journal = {Phys. Rev. A},
  volume = {76},
  issue = {1},
  pages = {012119},
  numpages = {6},
  year = {2007},
  month = {Jul},
  publisher = {American Physical Society},
  doi = {10.1103/PhysRevA.76.012119},
  url = {https://link.aps.org/doi/10.1103/PhysRevA.76.012119}
}

@article{arvidsson-shukur2024,
doi = {10.1088/1367-2630/ada05d},
url = {https://doi.org/10.1088/1367-2630/ada05d},
year = {2024},
month = {dec},
publisher = {IOP Publishing},
volume = {26},
number = {12},
pages = {121201},
author = {Arvidsson-Shukur, David R M and Braasch Jr, William F and De Bièvre, Stephan and Dressel, Justin and Jordan, Andrew N and Langrenez, Christopher and Lostaglio, Matteo and Lundeen, Jeff S and Halpern, Nicole Yunger},
title = {Properties and applications of the Kirkwood–Dirac distribution},
journal = {New J. Phys.}
}

@article{johansen2007_1,
title = {Reconstructing weak values without weak measurements},
journal = {Phys. Lett. A},
volume = {366},
number = {4},
pages = {374-376},
year = {2007},
issn = {0375-9601},
doi = {https://doi.org/10.1016/j.physleta.2007.02.039},
url = {https://www.sciencedirect.com/science/article/pii/S0375960107002654},
author = {Lars M. Johansen}
}

@article{hosten2008,
author = {Onur Hosten  and Paul Kwiat },
title = {Observation of the Spin Hall Effect of Light via Weak Measurements},
journal = {Science},
volume = {319},
number = {5864},
pages = {787-790},
year = {2008},
doi = {10.1126/science.1152697},
URL = {https://www.science.org/doi/abs/10.1126/science.1152697},
eprint = {https://www.science.org/doi/pdf/10.1126/science.1152697}}

@article{lundeen2011,
   title={Direct measurement of the quantum wavefunction},
   volume={474},
   ISSN={1476-4687},
   url={http://dx.doi.org/10.1038/nature10120},
   DOI={10.1038/nature10120},
   number={7350},
   journal={Nature},
   publisher={Springer Science and Business Media LLC},
   author={Lundeen, Jeff S. and Sutherland, Brandon and Patel, Aabid and Stewart, Corey and Bamber, Charles},
   year={2011},
   month=jun, pages={188–191}}

@article{danner2024_1,
author = {Danner, Armin and Masiello, Ismaele and Dvorak, Andreas and Kersten, Wenzel and Lemmel, Hartmut and Wagner, Richard and Hasegawa, Yuji},
year = {2024},
month = {10},
pages = {},
title = {Simultaneous path weak-measurements in neutron interferometry},
volume = {14},
journal = {Sci. Rep.},
doi = {10.1038/s41598-024-76167-6}
}

@article{pusey2014,
  title = {Anomalous Weak Values Are Proofs of Contextuality},
  author = {Pusey, Matthew F.},
  journal = {Phys. Rev. Lett.},
  volume = {113},
  issue = {20},
  pages = {200401},
  numpages = {5},
  year = {2014},
  month = {Nov},
  publisher = {American Physical Society},
  doi = {10.1103/PhysRevLett.113.200401},
  url = {https://link.aps.org/doi/10.1103/PhysRevLett.113.200401}
}

@article{kunjwal2019,
  title = {Anomalous weak values and contextuality: Robustness, tightness, and imaginary parts},
  author = {Kunjwal, Ravi and Lostaglio, Matteo and Pusey, Matthew F.},
  journal = {Phys. Rev. A},
  volume = {100},
  issue = {4},
  pages = {042116},
  numpages = {19},
  year = {2019},
  month = {Oct},
  publisher = {American Physical Society},
  doi = {10.1103/PhysRevA.100.042116},
  url = {https://link.aps.org/doi/10.1103/PhysRevA.100.042116}
}

@article{ritchie1991,
  title = {Realization of a measurement of a ``weak value''},
  author = {Ritchie, N. W. M. and Story, J. G. and Hulet, Randall G.},
  journal = {Phys. Rev. Lett.},
  volume = {66},
  issue = {9},
  pages = {1107--1110},
  numpages = {0},
  year = {1991},
  month = {Mar},
  publisher = {American Physical Society},
  doi = {10.1103/PhysRevLett.66.1107},
  url = {https://link.aps.org/doi/10.1103/PhysRevLett.66.1107}
}

@book{rauch2015,
  title={Neutron interferometry: lessons in experimental quantum mechanics, wave-particle duality, and entanglement},
  author={Rauch, Helmut and Werner, Samuel A},
  volume={12},
  year={2015},
  publisher={Oxford University Press}
}

@article{sponar2021,
  title={Tests of fundamental quantum mechanics and dark interactions with low-energy neutrons},
  author={Sponar, Stephan and Sedmik, Ren{\'e} IP and Pitschmann, Mario and Abele, Hartmut and Hasegawa, Yuji},
  journal={Nat. Rev. Phys.},
  volume={3},
  number={5},
  pages={309--327},
  year={2021},
  doi={https://doi.org/10.1038/s42254-021-00298-2}
}

@Article{danner2023,
title = {Neutron Interferometer Experiments Studying Fundamental Features of Quantum Mechanics},
author = {Danner, Armin and Lemmel, Hartmut and Wagner, Richard and Sponar, Stephan and Hasegawa, Yuji},
journal = {Atoms},
volume = {11},
year = {2023},
number = {6},
article-number = {98},
url = {https://www.mdpi.com/2218-2004/11/6/98},
issn = {2218-2004},
doi = {10.3390/atoms11060098}
}

@article{klepp2014,
    author = {Klepp, Jürgen and Sponar, Stephan and Hasegawa, Yuji},
    title = {Fundamental phenomena of quantum mechanics explored with neutron interferometers},
    journal = {Prog. Theor. Exp. Phys.},
    volume = {2014},
    number = {8},
    pages = {082A01},
    year = {2014},
    month = {08},
    issn = {2050-3911},
    doi = {10.1093/ptep/ptu085},
    url = {https://doi.org/10.1093/ptep/ptu085}
}

@article{hasegawa2003,
  title={Violation of a Bell-like inequality in single-neutron interferometry},
  author={Hasegawa, Yuji and Loidl, Rudolf and Badurek, Gerald and Baron, Matthias and Rauch, Helmut},
  journal={Nature},
  volume={425},
  number={6953},
  pages={45--48},
  year={2003},
  publisher={Nature Publishing Group UK London}
}

@article{summhammer1987,
  title = {Stochastic and deterministic absorption in neutron-interference experiments},
  author = {Summhammer, J. and Rauch, H. and Tuppinger, D.},
  journal = {Phys. Rev. A},
  volume = {36},
  issue = {9},
  pages = {4447--4455},
  numpages = {0},
  year = {1987},
  month = {Nov},
  publisher = {American Physical Society},
  doi = {10.1103/PhysRevA.36.4447},
  url = {https://link.aps.org/doi/10.1103/PhysRevA.36.4447}
}

@article{durstberger-rennhofer2011,
title = {Energy entanglement in neutron interferometry},
journal = {Physica B: Condensed Matter},
volume = {406},
number = {12},
pages = {2373-2376},
year = {2011},
issn = {0921-4526},
doi = {https://doi.org/10.1016/j.physb.2010.11.056},
url = {https://www.sciencedirect.com/science/article/pii/S092145261001121X},
author = {Durstberger-Rennhofer, Katharina and Hasegawa, Yuji}
}

@article{spekkens2005,
  title = {Contextuality for preparations, transformations, and unsharp measurements},
  author = {Spekkens, R. W.},
  journal = {Phys. Rev. A},
  volume = {71},
  issue = {5},
  pages = {052108},
  numpages = {17},
  year = {2005},
  month = {May},
  publisher = {American Physical Society},
  doi = {10.1103/PhysRevA.71.052108},
  url = {https://link.aps.org/doi/10.1103/PhysRevA.71.052108}
}

@article{catani2023,
  title = {Aspects of the phenomenology of interference that are genuinely nonclassical},
  author = {Catani, Lorenzo and Leifer, Matthew and Scala, Giovanni and Schmid, David and Spekkens, Robert W.},
  journal = {Phys. Rev. A},
  volume = {108},
  issue = {2},
  pages = {022207},
  numpages = {11},
  year = {2023},
  month = {Aug},
  publisher = {American Physical Society},
  doi = {10.1103/PhysRevA.108.022207},
  url = {https://link.aps.org/doi/10.1103/PhysRevA.108.022207}
}

@article{hornberger2012,
  title = {Colloquium: Quantum interference of clusters and molecules},
  author = {Hornberger, Klaus and Gerlich, Stefan and Haslinger, Philipp and Nimmrichter, Stefan and Arndt, Markus},
  journal = {Rev. Mod. Phys.},
  volume = {84},
  issue = {1},
  pages = {157--173},
  numpages = {0},
  year = {2012},
  month = {Feb},
  publisher = {American Physical Society},
  doi = {10.1103/RevModPhys.84.157},
  url = {https://link.aps.org/doi/10.1103/RevModPhys.84.157}
}

@article{waegell2017,
  title = {Confined contextuality in neutron interferometry: Observing the quantum pigeonhole effect},
  author = {Waegell, Mordecai and Denkmayr, Tobias and Geppert, Hermann and Ebner, David and Jenke, Tobias and Hasegawa, Yuji and Sponar, Stephan and Dressel, Justin and Tollaksen, Jeff},
  journal = {Phys. Rev. A},
  volume = {96},
  issue = {5},
  pages = {052131},
  numpages = {8},
  year = {2017},
  month = {Nov},
  publisher = {American Physical Society},
  doi = {10.1103/PhysRevA.96.052131},
  url = {https://link.aps.org/doi/10.1103/PhysRevA.96.052131}
}

@article{wagner2023,
  title={Quantum causality emerging in a delayed-choice quantum Cheshire Cat experiment with neutrons},
  author={Wagner, Richard and Kersten, Wenzel and Lemmel, Hartmut and Sponar, Stephan and Hasegawa, Yuji},
  journal={Sci. Rep.},
  volume={13},
  number={1},
  pages={3865},
  year={2023},
  publisher={Nature Publishing Group UK London}
}
\section*{Acknowledgments}
This research was funded in part by the Austrian Science Fund (FWF) (grant DOI: 10.55776/P34105). For open access purposes, the authors have applied a CC BY public copyright license to any authors accepted manuscript version arising from this submission. The authors acknowledge the hospitality of ILL. The authors acknowledge TU Wien Bibliothek for financial support through its Open Access Funding Program. Y.H. was partly supported by JSPS KAKENHI Grant No. JP25K03386. We thank Tom Rivlin for the useful discussions.
\section*{Author contributions}
I. V. M.: theoretical formulation, experimental implementation, data analysis, manuscript writing. H.L.: experimental design and implementation. A.D.: experimental implementation. S.S.: manuscript writing and 3D setup figure. Y.H.: supervision, theoretical formulation, experimental design and implementation, manuscript writing. All authors have revised and accepted the manuscript.
\section*{Competing interests}
The authors declare no conflicts of interest.

\end{document}